\documentclass[10pt,journal]{IEEEtran}

\usepackage{amsfonts,amssymb,amscd,xspace,mathrsfs}
\usepackage[fleqn]{amsmath}
\usepackage{graphics,graphicx,epsfig,url}
\usepackage{array,pslatex}

\newtheorem{theorem}{Theorem}

\newtheorem{definition}{Definition}

\newcommand{\citet}[1]{\citeauthor{#1}~\shortcite{#1}}

\let\originaleqref\eqref
\renewcommand{\eqref}{Equation~\originaleqref}

\newcommand{\I}{\ensuremath{\mathscr{I}}}

\renewcommand{\H}{\ensuremath{\mathscr{H}}}

\renewcommand{\Re}{\ensuremath{{\mathbb R}}}

\renewcommand{\d}{\ensuremath{{\mathbf d}}}
\newcommand{\p}{\ensuremath{{\mathbf p}}}
\newcommand{\x}{\ensuremath{{\mathbf x}}}
\newcommand{\one}{\ensuremath{{\mathbf 1}}}
\newcommand{\RevDef}{\ensuremath{\mathrm{RD}}}
\newcommand{\Trans}{\ensuremath{^\top}}

\newcommand{\reg}{\textsuperscript{\textregistered}}

\usepackage[]{color}
\def\arch{\textcolor{black}}

\begin{document}
%
\title{An Iterative On--Line Mechanism for\\ Demand--Side Aggregation}
%
%
%
%

\author{Achie~C.~Chapman 
and 	Gregor~Verbi\v{c}
\IEEEcompsocitemizethanks{\IEEEcompsocthanksitem A. Chapman and G. Verbic are with the 
School of Electrical and Information Engineering, University of Sydney, NSW, Australia, 2006.\protect\\
E-mail: archie.chapman@sydney.edu.au}
}

\maketitle

\IEEEdisplaynontitleabstractindextext

%
\IEEEpeerreviewmaketitle

\begin{abstract}
This paper considers a demand--side aggregation scheme specifically for large numbers of small loads, 
such as households and small and medium--sized businesses.
We introduce a novel auction format, called a \emph{staggered clock--proxy auction} (SCPA),
for on--line scheduling of these loads.
This is a two phase format, consisting of:
a sequence of overlapping iterative ascending price \emph{clock auctions}, 
one for each time--slot over a finite decision horizon, and; 
a set of \emph{proxy auctions} that begin at the termination of each individual clock auction, 
and which determine the final price and allocation for each time--slot.
The overlapping design of the clock phases grant bidders the ability to effectively bid on inter--temporal bundles of electricity use, 
thereby focusing on the most--relevant parts of the price--quantity space. 
Since electricity is a divisible good, the proxy auction uses demand--schedule bids, 
which the aggregator uses to compute a uniform--price partial competitive equilibrium for each time slot.
We show that, under mild assumptions on the bidders' utilities functions, 
the proxy phase implements the Vickrey--Clarke--Groves outcome, 
which makes straightforward bidding in the proxy phase a Bayes--Nash equilibrium.
Furthermore, we demonstrate the SCPA in a scenario comprised of household agents with three different utility function types, 
and show how the mechanism enables efficient on--line energy use scheduling.
\end{abstract}


\section{Introduction}\label{sec:Introduction}
\noindent
\IEEEPARstart{D}{emand} response refers to methods for influencing the amount of power drawn from an electrical power system by end--users, 
thereby making electrical loads a resource that can be used to undertake control actions, 
such as load--balancing, peak load shaving, emergency load shedding and ancillary services. 
Demand response is employed to provide additional capacity to the power system without costly new infrastructure,
and to facilitate greater penetration of renewable generation,
as increasingly flexible energy use is able to better track the intermittent supply provided by 
many renewable sources of energy. 
In this paper, we develop a \emph{small--load demand--response aggregation} (SL--DR) scheme;
that is, a scheme constructed specifically for aggregating the large numbers of small loads, 
such as households and small-- and medium--sized enterprises,
spread across an electrical distribution network.

We adopt a typical framework comprising an SL--DR \emph{aggregator} that coordinates, 
schedules or otherwise controls part of participating electrical loads.
\arch{ The aggregator may be a retailer trying to keep energy purchasing costs down by encouraging electricity use at cheaper times, 
or a third party trading price differences on a wholesale electricity market. 
We assume that the aggregator operates in either a highly competitive environment, 
or that its prices/margins are regulated, 
so that it is only able to extract a small charge per kWh from its customers.  
In effect, this is the same as assuming that the aggregator is benevolent, 
and only passes on its cost of purchasing energy in the wholesale market.  
On the customer side, we assume that each user has (at least partial) automation of a proportion of its interruptable and deferable loads 
by employing an \emph{energy management system} (EMS), 
which controls or schedules devices include a hot water storage system, home ventilation and air conditioning, dish washers, clothes washers and dryers, and so on.  
Indeed, the scheme developed here might only be applied to selected devices, which are put under the EMS's control, 
and for which the customer has no want or need to draw power from the grid at a specific time.}

In particular, this paper investigates an auction--based aggregation scheme for 
scheduling energy use in a future power system.
In doing so,
we directly confront three major difficulties inherent in aggregating residential loads.
First, we note that most small users, such as households, 
have a demand for electrical energy this is inherently \emph{combinatorial} and \emph{non--convex} in structure. 
\arch{
Specifically, many small user's electrical loads possesses both inter--temporal complementarity and substitute effects.
Substitute effects primarily involve shifting the timing that a device draws a load from the grid 
(e.g.~actively controlling the recharge schedule of a hot water storage system or the start time of a dishwasher).
An example of a complementarity is a residential electricity user that has a need to both wash and dry their clothes; 
however, washing must be done before drying can start.
In addition to complementarities,
other non--convexities arise when users' devices have discrete operating points, 
such as heating and cooling devices that use compressors, or washing and cleaning devices with set programs.}
Second, users' preferences over energy use patterns are \emph{private},
and are unknown to the aggregator (or any other user).
\arch{ As such, any SL--DR scheme needs to consider the difficulties of implementing an efficient allocation 
when users can misreport their preferences.}
Third, considering the large number of participants required to make an SL--DR scheme financially viable, 
\arch{ any system comprising an aggregator and its users will face significant communication and computation requirements.}

To this end, we introduce a novel combinatorial clock--proxy auction format~\cite{AusubelCM2006} 
tailored to the on--line electricity use scheduling problem.
Existing clock--proxy auction formats only consider static problems, 
such as radio spectrum allocation,
and so are not directly applicable to the dynamic, on--line environment of SL--DR.
Thus, the main contribution of this paper is to adapt the techniques of clock--proxy auctions 
to the problem of on--line scheduling of a resource 
with users that have preferences defined over combinations of different times of resource use.

In more detail, we propose a \emph{staggered clock--proxy auction} (SCPA), 
an iterative auction format consisting of a sequence of overlapping clock auctions, 
one for each time--slot in a day, extending out to a suitable finite decision horizon (i.e.~one day), 
with the final allocation and prices in an individual time--slot determined by a proxy auction.
\arch{ 
In a practical implementation, the SCPA's auction slots are expected to align with the local energy market's operation.  
For example, in the Australian National Electricity Market, supply procurement auctions are run for each 30 minute period;  
the SCPA is constructed in such a way that iteration can be completed within this duration.  
Moreover, the SCPA's operation is both: 
(i) oblivious to the preference models held by the users or their EMS agents, 
and as such, it is not tied to any one particular preference representation or utility model, and 
(ii) any monotonic cost function can be employed by the aggregator, 
including those generated by the piece--wise linear supply functions used in many wholesale electricity markets.
This makes the SCPA flexible and generally applicable, 
which stands in contrast to many proposed aggregation mechanisms that are tied to specific utility and/or cost models 
(see Section~\ref{sec:RelatedWork} for a detailed comparison). }

SCPAs retain three important features of combinatorial clock--proxy auctions for static settings:
(i) 	anonymous (uniform) linear prices;
(ii) 	monotonic price changes; and
(iii)	activity rules for quantity changes.
By preserving these features, 
SCPAs directly addresses the three difficulties facing aggregators of residential loads listed above.
First, 
bidders 
can in effect bid on inter-temporal bundles of electricity use,
by adjusting their bids across all time--slots according to all prices for all time--slots,
thereby matching their potentially complex preferences for electricity use.
Second, 
incentives for participants to ``game'' the aggregation scheme, 
by interacting strategically rather than sincerely,
are reduced through the SCPA's use of well-designed price-- and bid--adjustment rules in the clock phase,
and by implementing competitive equilibrium prices 
in the proxy phase.
Third, the iterative nature of SCPAs reduces the communication and computation requirements of the aggregator and bidders, 
\arch{by focusing participants' bids on the most relevant parts of the price-quantity space in the clock phase, 
and using this to restrict the price interval over which demand levels are specified in the proxy phase.
This compares favourably to direct auction mechanisms, 
which typically rely on solving centralised winner-determination and cost division problems.}

Furthermore, our SCPA rolling horizon format is appropriate in the SL--DR problem, 
given the recurring nature of many households' tasks and energy use requirements 
(e.g.~space heating and cooling, hot water storage, cleaning, and so on), 
\arch{as these tasks do not require planning over a horizon of longer than a day.}
In addition to the above, 
we expect that users know only some broad energy requirements at long horizons, 
and that more detailed schedules are available only as the time of use draws near.
Thus, an on--line mechanism with an overlapping initial phase and staggered closing phase is appropriate for this setting.

\arch{Finally, it should be noted that by using a dynamic pricing mechanism such as the SCPA, 
an aggregator is able to pass some price volatility risk on to the end user. 
This sharing of risk is complemented by our assumption that users have some automated devices that can respond to dynamic pricing.  
A user may also limit its exposure to varying prices by only including some devices in the scheme, 
e.g.~to only include hot water recharging and the start time of the dishwasher overnight.  
In this way, the customer can choose their level of exposure to the risk inherent in time--varying prices.  
That said, one of the key benefits of deploying schemes such as the SCPA at a large scale is that they are expected to reduce price variability, 
as flexible loads are shifted to time--slots with lower prices and away from those with higher prices.  }

\begin{figure}
 \centering
 \includegraphics[width=0.40\textwidth]{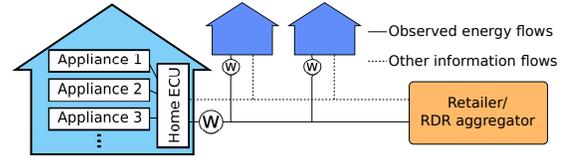}
 \caption{ Aggregator and users interaction:
 control of appliances remains completely under a user's control, via the Home ECU;
 the aggregator observes energy flows via the metering infrastructure (solid lines);
 other information flows are facilitated via a communications infrastructure (dashed lines).
 }
 \label{fig:DetailedInteractionModel}
\end{figure}

In summary, the contributions of this paper are as follows:
\begin{enumerate}
 \item We develop an iterative on--line auction protocol, SCPAs, appropriate for small-load electricity-use scheduling;
 \item We include costly supply into the analysis of clock--proxy auctions 
 (c.f.~standard clock--proxy auctions where the seller is assumed to have no value for the goods);
 \item For this more general setting, we prove that straightforward bidding is an equilibrium of an individual proxy phase of SCPAs; and
 \item We demonstrate a specific implementation of a SCPA for an energy use scheduling problem 
	comprising households with three different utility representations, 
	\arch{which shows that the method is feasible for most standard duration time--slots for a large number of users,
	and that the SCPA can help reduce price variability in wholesale markets, 
	thereby reducing total system energy costs.}
\end{enumerate}

The paper progresses as follows:
The next section reviews the literature on demand response, with focus on applications of online mechanism design.
Section~\ref{sec:ModelProblemDesc} introduces the model and describes the energy use scheduling problem addressed.
In Section~\ref{sec:SCPA}, we define the general SCPA format, including two different bid--adjustment rules, 
and in Section~\ref{sec:Analysis}, 
we analyse the general properties of SCPAs, prove some important equilibrium results for the proxy phase, 
and discuss the sequence of prices and allocations that the SCPA produces.
Section~\ref{sec:Demonstration} describes and evaluates a specific implementation of a SCPA for an energy use scheduling problem.
Finally, Section~\ref{sec:Conclusions} summarises and lists future directions.


\section{Related Work} 
\label{sec:RelatedWork}\noindent
This section contains a brief review of related approaches to online SL--DR problems.
A general survey of non-cooperative games applied to demand--response is presented in~\cite{SaadEtal2012},
and a detailed critique can be found in~\cite{ChapmanVH_IREP2013}.

Several recent works 
adopt a mechanism design approach to energy use scheduling.
Both~\cite{SamadiEtal2012} and~\cite{SteinEtal2012} propose Vickrey--Clarke--Groves (VCG) based mechanisms,
in line with those derived in~\cite{ParkesEtal2010,BergemannValimaki2010}; and
another, related approach is derived in~\cite{TanakaCL2012}.
However, VCG is a direct mechanism, so to use it, the aggregator can take one of two approaches.
First, the aggregator could ask for valuations to be reported over a complete set of demand bundles,
but in this case the size of the required preference representations grows exponentially as the model becomes more fine-grained 
(i.e.~considers shorter and more time--slots).
Second, the aggregator could enforce a compact preference representation. 
For example, \cite{SamadiEtal2012} use a simplified preference model in the form of a convex function taking a small number of parameters, 
while~\cite{SteinEtal2012} achieve the same by considering only one device, a plug--in electric vehicle.
Nonetheless, the underlying optimisation problem used to compute VCG allocations and payments grows exponentially 
in the number of participants and with the size of their messages.
Both of these effects have severe detrimental effects on the feasibility of communication to and computation by the aggregator.
Thus, we consider an iterative auction design.

Several iterative mechanisms are related to our work. 
The authors of~\cite{IwasakiEtal2013} consider an iterative combinatorial auction for a collection of divisible goods. 
Their technique implements an efficient allocation by individually querying bidders about their types and 
ensuring sincere responses by implementing transfers that are the same as those for VCG in expectation.
The techniques used in that paper cannot be directly applied to our setting with divisible goods and a large number of bidders, 
but both make use of iterative ascending prices to implement Pareto efficient allocations.
An ascending-price method similar to the SCPAs clock phase is proposed in~\cite{VeitEtal2013} for energy consumption scheduling. 
The proposed method is proven to converge, but is not analysed with respect to the strategic opportunities available to energy users.

Finally, several auction formats currently in use are closely related to the SCPA.
The SCPAs proxy phase is reminiscent of the reverse supply function auctions 
used in wholesale electricity markets for procuring generation capacity \cite{BaldickGK2004}, 
and also auctions of US Treasury Securities \cite{WangZender2002}.
Electricit\'{e} de France (EDF) until recently used a clock auction format in their generation capacity auctions \cite{AusubelCramton2010}.
Various authorities around the world have used combinatorial clock--proxy auctions to sell radio-frequency spectrum, 
e.g.~the Australian Communications and Media Authority's 700 MHz and 2.5 GHz spectrum auctions \cite{AusubelBaranov2014}.


\section{Model and Problem Description} 
\label{sec:ModelProblemDesc} \noindent
Throughout, the set of real numbers is denoted $\Re$, 
and $\one$ is an all-ones vector of length given by context.
We adopt a discrete-time model, where operations are divided into 
$\H = \{t,t+1,\ldots,t+H-1\}$, consisting of $H$ time--slots over the decision horizon.
Consequently, all electrical quantities are stated as blocks of energy; 
for example a 100W appliance running for $30$ minutes is described by a $0.05$kWh demand block.

The on---line energy--use scheduling model consists of:
\begin{itemize}
	\item A divisible good, electricity, with supply level in time--slot $h$ of $x_h\in\Re_+$;
	\item A set of agents $\I = \{0,1,2,\ldots,I\}$, where $0$ is the \emph{aggregator}, with: 
	\begin{itemize}
	 \item 	A set of cost functions $c_h:\Re_+\to\Re_+$, one for each time slot to capture varying costs of generation,
	 where $c_h(x_h)$ is the cost to the aggregator of supplying $x_h$ units of electrical energy in time--slot $h$, 
	 (it has no further intrinsic value to the aggregator);
	\end{itemize}
	and each $i\neq0$ is an electricity \emph{user} agent, with:
	\begin{itemize}
	  \item A level of demand for electrical energy, $d^i_h\in \Re_+$ during time--slot $h \in \H$, 
		such that $x_h = \sum_{i\in\I\setminus 0} d^i_h$;
	  \item A preference function over levels of electricity use in each time--slot,
	  $v^i: \Re^{H}_+ \to \Re_+$, 
	  where $v^i(\d^i,\theta^i)$ is the value to $i$ for a demand profile over time of $\d^i = [d^i_{t}, d^i_{t+1},\ldots,d^i_{t+H-1}]\Trans$ 
	  when its private state (type) is given by an information structure $\theta^i$.
	\end{itemize}
\end{itemize}

\arch{In this setting, the information structure $\theta^i\in\Theta$ represents the state of a set of tasks or activities
that the user employs electrical energy to complete.
As such, users' preference functions should accommodate both complementary and substitution effects 
in power use across the different time slots.
These effects are the result of electricity's use as an \emph{intermediate good}: 
electricity is not consumed per se, 
but put to use to perform tasks such as heating and cooling, cooking, cleaning, lighting and entertainment. 
The patterns of demand for completion of these tasks ---
which may have inter--temporal complementarities, order relations, or might be substituted between different times --- 
is then manifest in complicated preferences for electricity use.}

Given this general model, the SL--DR problem is to derive the method by which an aggregator optimally structures its interaction with users.
That is, an SL--DR scheme defines how the aggregator divides the costs it faces among its users, 
and induces them to use electricity at the most appropriate times. 
\arch{Let $\d = [\d^1, \ldots, \d^I]$ be the concatenation of electricity use vectors over the horizon;
the profile of demands for time--slot $h$ is its $h^{th}$ row, denoted $\d_h$.}
In general, we can define the cost division used by the aggregator as a vector function, 
$\phi(c_h(x_h),\d_h)$,
which returns a vector of costs, one for each user. 
In this, the aggregator calculates 
$x_h = \d_h\cdot\one$.

Given the users' preferences and costs above, 
we can define a \emph{utility function} $u^i_h(\d)$, 
which takes a quasi--linear form (linear in prices), to combine their values and costs for using electricity:
\begin{equation}\label{eqn:totalUtility}
 u^i(\d) = v^i(\d^i,\theta^i) - \sum_{h=t}^{t+H-1} \phi^i(c_h(x_h),\d_h),
\end{equation}
where $\phi^i(c_h(x_h),\d_h)$ is the $i^{th}$ component of the aggregator's cost division function.
\eqref{eqn:totalUtility} shows that the aggregator and users' actions are coupled 
through their dependence of their utilities on the vector of total loads over \H, $\x = [x_{t},\ldots,x_{t+H-1}]$.
Thus, their interaction results in a game; 
and since each user's state, $\theta^i$, and therefore its reward function, $v^i$, is private, 
it is a game of \emph{incomplete information}.

\arch{
A standard solution to these games is given by the Bayes-Nash equilibrium condition.
\begin{definition}
A Bayes-Nash equilibrium is a set of demands $\d^i$ for the $N$
competitive bidders such that for each bidder $i$ , $\d^i$ maximises its expected
utility of profit for all $\theta^{-i}\in\Theta^{-i}$.
\end{definition}
}

\arch{
The desirable properties that an auction may exhibit include 
allocative efficiency, social welfare maximisation, incentive compatibility and budget balancedness. 
An allocation is \emph{efficient} if the goods go to the bidders with the highest valuations. 
This corresponds to the notion of \emph{social welfare} maximisation, 
where an allocation is efficient if it maximises the sum of all agents' utilities; in our setting this is given by:
\begin{equation*}
\sum_{i\in\I} u^i(\d^*) \geq \sum_{i\in\I} u^i(\d) \quad \forall\ \d
\end{equation*}
A auction mechanism is called \emph{incentive compatible} 
if all agents do their best by {truthfully} reporting their private information to the auctioneer.
A mechanism can either be either \emph{dominant strategy incentive compatible}, 
in which case the best action of any bidder is to truthfully report regardless of what others do;
or it can be \emph{Bayes-Nash incentive compatible}, in which case truth--telling forms a Bayes-Nash equilibrium, 
as defined above.
Finally, a mechanism is called (ex-post) \emph{balanced} if it does not require money to be injected or withdrawn 
to balance the payments between participants
(cf.~some mechanisms implement efficient outcomes but at a risk of running a deficit).
}

\section{The SCPA Auction Format} 
\label{sec:SCPA}\noindent
In this section, we describe the generic SCPA format, while Section~\ref{sec:Analysis} provides analysis of the mechanism.

The SCPA is an iterative simultaneous auction, 
consisting of $H$ live slot auctions, 
one for each time slot $h\in\H$. 
The cost division function is defined for each slot independently, 
and is \emph{linear} and \emph{anonymous}, 
so that each slot has one price $p^h$ and costs are proportional to use, and as  such, prices do not depend on the identity of the buyer; 
that is, $\forall h\in\H$: 
\begin{equation*}
  \phi_h(c_h(x_h),\d) = p_h \cdot \d_h.
\end{equation*}
This stands in contrast to the often-used VCG mechanism, 
which uses both \emph{nonlinear} and \emph{discriminatory pricing}.

For each time--slot, the SCPA consists of two phases: 
(i) an ascending-price clock phase, which runs in parallel with the clock phases of other time--slots in \H, and 
(ii) an individual proxy phase, with the closing times of these phases staggered in order of their time--slot.
All $H$ clock phases are run together, 
but are paused when the next-closing time--slot $t$'s proxy auction is run.
Thus, the procedure alternates between a clock auction phase, 
during which bids are placed on all $h\in\H$,
and a proxy auction phase, 
during which bids are placed for allocations in the next-closing time--slot, $t$, 
while the other time--slots' prices are held fixed.
The final allocation and unit price for time--slot $t$ is determined in the proxy phase, 
after which the entire procedure moves forward one time--slot, 
and a new clock auction for time--slot $t+H$ is opened.
In total, each time--slot is involved in $H$ clock phases before its final proxy phase, 
punctuated by $H-1$ proxy phases for the preceding live time--slots.
All bids placed during the clock phases are considered live in the subsequent proxy auction phase.
Bids in the clock phases are also subject to bidding and activity rules, 
some of which are carried over into the proxy phase, 
thereby binding bidders' proxy-phase bids to their bids in earlier clock phases.

\begin{figure}
 \centering
 \includegraphics[width=0.38\textwidth]{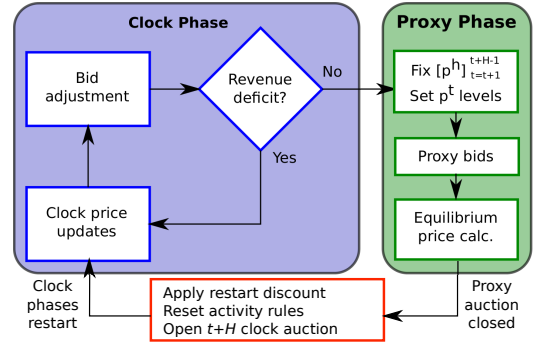}
 \caption{Schematic of the SCPA, showing separate clock and proxy phases. }
 \label{fig:SCPAuction}
\end{figure}

The general SCPA procedure is outlined in Fig.~\ref{fig:SCPAuction}. 
The next two sections provide the details of the clock and proxy phases. 


\subsection{Clock Phase}
Within each time--slot $h$, there is a sequence of $k\geq 1$ clock auction iterations.
Each iteration consists of a {price adjustment} step 
and a {bid update} step, 
as indicated by the while loop in the left box in Fig.~\ref{fig:SCPAuction}.
In this process: 
(i) the aggregator updates prices in response to bids, depending on a measure of \emph{revenue deficit} ($\RevDef$),
then 
(ii) the buyers update their bids in response to the new prices, according to their own preferences, 
while 
(iii) iterations stop when the process meets a termination condition that also depends on the level of excess demand. 

Details of these three steps are given in the sections below, 
but before beginning, we introduce some notation:
at iteration $k$ for all active time--slot auctions $h\in \H$, 
let $\p^k$ be the $H\times 1$ vector of prices.
For each agent, $i\in\I$, let $^i\d^k$ be the $H\times 1$ vector of its bids.

\subsubsection{Price Adjustment}
\label{sssec:PriceAdjustment}\noindent
In each time--slot auction, price adjustment is monotonic, 
with prices beginning low and rising until the revenue deficit in that time--slot is (close to) eliminated.
For each slot $h\in\H$, let the per-unit revenue deficit when $x^k_h$ units of energy are supplied be given by:
\begin{equation}\label{eqn:RevDef}
 \RevDef^k_h = c(x^k_h)/x^k_h - p^k_h.
\end{equation}
In this expression, the term $c(x^k_h)/x^k_h$ is \emph{average total cost} (ATC), 
and it plays an important role in the proxy phase.

In the first clock auction for time--slot $t+H$, 
prices are initialised a point $p^0_h$ that is certain to be less the aggregator's average total cost.
This can be $p^0_h =0$, which is reasonable because total energy demand is bounded,
but higher will result in fewer clock--phase iterations.
\arch{
In subsequent restarts, 
the price of electricity in a time--slot begins close to its closing price in the previous clock phase.
In practice, 
a small discount is applied to the closing prices to accommodate changes in preferences flowing from unforseen random events, 
i.e.~to allow for uncertainties in energy use and generation costs.
}

Within a single clock auction phase, let $x^k_h$ be the total demand in time--slot $h$ at iteration $k$, 
and $p^k_h$ be the price inducing that level of demand.
If $\RevDef >0$, 
such that the price of supplying $x^k_h$ units of energy is not covered by the auction's revenue, 
then the price increases. 
Otherwise, the price stays the same in the next iteration $k+1$; that is, 
the price for each slot is adjusted by:
\begin{equation}\label{eqn:PriceUpdate}
  p^{k+1}_h = 
  \begin{cases}
    p^k_h +\delta^{k+1}  & \mbox{if } \RevDef^k_h > 0,\\
    p^k_h     & \mbox{otherwise.}
  \end{cases}
\end{equation}
where $\delta^{k+1}$ is a configurable price adjustment step-size.

\subsubsection{Bid Updates}
\label{sssec:BidUpdates}
\noindent
As in standard combinatorial clock auctions,
bid quantity adjustments are constrained by bidding and activity rules, 
which demarcate the space of admissible bid changes in response to price adjustments.
%
%
The activity rule we consider is based on the microeconomic principle of \emph{revealed preferences}, 
as introduced for clock--proxy auctions in \cite{AusubelCM2006}.	
\begin{definition} 
\emph{Revealed-preference (RP) bid constraint}.\ 
For all iterations $k,l\geq1$, $k<l$, an agent's bid across the decision horizon must satisfy: 
\begin{equation}\label{eqn:rpc1}  
  (\p^k)^\top \cdot ^i\d^k - (\p^k)^\top \cdot ^i\d^l \geq 0, 
\end{equation}
and:
\begin{equation}\label{eqn:rpc2}  
  (\p^l)^\top \cdot ^i\d^l - (\p^l)^\top \cdot ^i\d^k \geq 0.
\end{equation}
\end{definition}
The two constraints above imply that bundle $^i\d^k$ is preferred to $^i\d^l$ under prices $\p^k$,
because the agent is willing to pay more in total for $^i\d^k$,
and vice versa for prices $\p^l$.
We require that these two constraints are satisfied for every bid in every iteration of the clock phase, 
but that they are reset when the clock auction is restarted after a proxy phase closes one auction.


This rule deters the users from gaming the system.
\arch{
For example, 
a group of users may try to systematically over--bid for energy in some slots in order to drive other users away, 
then reduce demand in the clock--phase after those others have purchased more energy in earlier time--slots (this is often called bid \emph{parking}).
The opportunities to profit from this and other styles of gaming are mitigated by the RP bid constraints, 
but we emphasise that to be effective, the RP bid constraints must be also applied to bids placed in the proxy phase;
indeed,~\eqref{eqn:rpc1} and~\eqref{eqn:rpc2} are the link between the clock and proxy phases.
If they are not carried over, then the bidders are not bound to the prices discovered in the clock phase. 
In contrast, under the price adjustment and restart rules described above, 
the effects of the RP bid constraints are carried across phases of the SCPA.
}

%


\subsubsection{Termination Condition}
\label{sssec:Termination}\noindent
In each clock auction, prices stop ascending if $\RevDef_{h} \leq 0$.
Note that demand for energy in $h$ can rise when prices in other time--slots from which energy can be substituted increase, 
so that $\RevDef_{h}$ again becomes $>0$, and the price recommences ascending.
However, the entire clock phase terminates when $\RevDef_{h} <0$ for all \H.

%

\subsection{Proxy Phase}
\noindent
The proxy phase operates only for time--slot $t$, the next--closing time--slot's allocation. 
During the proxy phase for $t$, all other time--slots' prices are fixed at their last clock phase levels $\p_{\H\setminus t}$.
In the proxy phase, users bid their demand levels for electricity over a restricted interval of prices, 
which the aggregator uses to compute a final price,
as illustrated in Fig.~\ref{fig:SCPAuction} and detailed below.
\subsubsection{Price Interval and Breakpoints}
\noindent
The first step in the proxy phase is for the aggregator to determine a price interval and breakpoints,
which are chosen to ensure that the user demand schedule generated from the users' bids intersects with the supply function.
Denote interval ends and breakpoints by 
$P = [P_0,P_1,\ldots,P_{|P|}]$.
%
We do not prescribe a procedure, 
but it suffices to say that relatively simple techniques can be used to ensure supply crosses demand. 
These might include analysing bids in the preceding clock phases, 
or the history of closing prices in previous proxy phase for this time--slot during the day,
and estimating the slope of the aggregate demand curve around the clearing prices 
or placing a confidence interval around a predicted clearing price. 
Breakpoints can then be spread uniformly between the interval ends, %
or their spacings can be scaled according to some distributional information about where the MCE price will fall.

\subsubsection{Demand Schedules}
Bidders pass a vector of demand levels to the aggregator: 
\begin{equation*}
 D^i(P) = [d^i(P_0),d^i(P_1),\ldots,d^i(P_{|P|})],
\end{equation*}
corresponding to the breakpoints $P$.
These marginal bid levels are subject to the RP bid constraints.
Additionally, the bidders' demands must satisfy: 
$d^i(P_l) \geq \d^i(P_m)$ for all $P_l < P_m$,
so that each bidder's demand curve is weakly downward-sloping.
\arch{
This reflects the typical microeconomic assumption that consumption of each unit more of a good has decreasing benefit.
In contrast to the clock phase, in time--slot $t$'s proxy phase, the other time--slot prices are fixed, 
so this assumption can be applied without having to consider flow-on effects from simultaneously increased prices in other time--slots.}

The aggregator computes the aggregate demand schedule $D$ from the demand levels $\{D^i(P)\}_{i\in \I\setminus 0}$ 
by first calculating: 
\par \vspace{-0.9\baselineskip} 
\small
\begin{equation*}
D = \left[ \sum_{i\in\I\setminus0} d^i(P_k) \right]_{k=0}^{|P|}
\end{equation*}
\normalsize
and then taking the linear interpolation of values in $D$ as the aggregate demand curve.

%

\subsubsection{Computing the Final Price and Allocation}
\noindent
The intersection of the aggregate demand schedule and the supply function determines the final price for energy in a time--slot, 
which the bidders use to choose their level of demand.



Specifically, given $D$, 
the aggregator solves for the intersection of ATC with the line segments between the breakpoints, 
which can be done numerically.
%
%
%
%
This approach may appear unsatisfying in the cases of small numbers of users, 
but for large numbers, we are confident that computing prices in this way will produce a valid solution, 
by appeal to the Shapley-Folkman theorem \cite{Starr1969_ShapleyFolkman}.
\arch{This is because the relative effects of discrete changes in electricity use (e.g. turning a device on or off) 
are much larger for fewer users than for larger numbers of users.  
In other words, the aggregate demand schedule approches a continuous curve as the number of users grows.}

\section{Analysis of SCPA} \label{sec:Analysis}
We now analyse the operation of SCPAs 
and discuss how they bring about an efficient allocation in on--line electricity use scheduling problems.
This analysis starts with the proxy phase and works backwards to the clock phase, 
as would be done to analyse a finite--horizon dynamic program.

\arch{
We begin by noting that the purpose of the clock phase of the SCPA is to aid price-discovery 
and facilitate energy use coordination between the users, 
not to fix an allocation or determine final prices.
The proxy phase is then used to find an efficient allocation at ``good'' prices for the next time--slot. 
Thus it is the proxy phase that receives the most detailed analysis, 
and for which we provide the strongest results. }

\subsection{Analysis of Individual Proxy Auctions}
\label{sec:PAanalysis}
In order to analyse the proxy phase, we now introduce some concepts from cooperative game theory.
In this section, the time--slot index $h$ is dropped, 
because only one proxy auction is run at a time, 
and the private information structure $\theta^i$ is omitted 
because the agents implicitly report this information to the aggregator in the form of their proxy bids.

A (static) transferable utility (TU) game is a pair $\langle \I,w\rangle$, 
where $\I$ is a set of players,
and $w(S)$ is a \emph{characteristic function}, $w:2^I\to \Re_+$ with $w(\emptyset) = 0$, 
that maps from each possible coalition $S$ to the \emph{worth} of $S$.

Call the proxy phase of the SCPA a \emph{demand--schedule bid} (DSB) game.
The function $w(S)$ defines the worth of $S\subseteq\I$ 
as the sum of all participants' values, which is the value of the electrical energy provided by the aggregator,
less the aggregator's costs (as long as the aggregator is included, otherwise no electricity is traded); 
that is:
\begin{equation} \label{eqn:Cvalue}
 w(S) = 
 \begin{cases}
 \displaystyle\max_{\d} \sum_{i\in S\setminus 0} v^i(d^i) - c\left( \sum_{i\in S\setminus 0} d^i \right) & \text{if } 0\in S \\
 0 & \text{otherwise.}
 \end{cases}
\end{equation}

In general, a \emph{payoff} in a TU game is a vector of utilities, $[u^i]_{i\in\I} \in \Re^I$.
We continue to assume that the users utilities are quasi-linear, 
as in~\eqref{eqn:totalUtility}.
Given this, changes to the cost division function $\phi$ alters the payoff profile by transferring utility among the market participants, 
but it does not alter the coalition's worth, 
\arch{as this is only given by the values and costs of using and supplying electricity within the aggregation scheme.}

The stable outcomes of the proxy auction are characterised using the concept of the \emph{core} of cooperative game.
\begin{definition}\label{def:core}
 \arch{The core of a TU game $\langle \I,w\rangle$ is the set of payoffs satisfying two conditions:
 (i) the payoffs must share the full worth of $\I$ among its members, and
 (ii) the payoffs must provide no opportunity for a profitable deviation to any subset of players $S\subset\I$.
 Formally, these conditions are expressed as:}
\par \vspace{-0.9\baselineskip} 
\small
\begin{equation*}
 Core(\I,w) = \left\{ [u^i]_{i\in\I} :\sum_{i\in\I} u^i = w(\I),\ \sum_{i\in S} u^i \geq w(S)\ \forall S\subseteq\I  \right\}
\end{equation*}
\normalsize
\end{definition}
%
Note that payoffs in the core are \emph{efficient}, by the definition of a coalition's worth function in~\eqref{eqn:Cvalue}, 
and also that no money is in injected into or removed from the coalition (i.e.~the transfers are \emph{balanced}).

It has long been known that in all TU \emph{exchange economy games} (a.k.a.~Edgeworth boxes) 
with convex increasing utility functions and fixed endowments of goods, 
the core has positive measure~\cite{ShapelyShubik1969}.
However, a DSB game with costly production (i.e.~by the aggregator) needs to satisfy an additional condition:
that the slope of the aggregator's cost function 
is less than the slope of the sum (or aggregate) of the bidders' utility functions at zero allocation,  
or in microeconomic terminology, that the market supply function is less than the demand function at zero demand.
This ensures that average total cost and demand intersect at a positive price and level of supply. 
Assuming demand is weakly downward sloping, 
then this is readily satisfied in our setting,
so the DSB game has a core.

\begin{figure}
	\centering
	\includegraphics[width=0.26\textwidth]{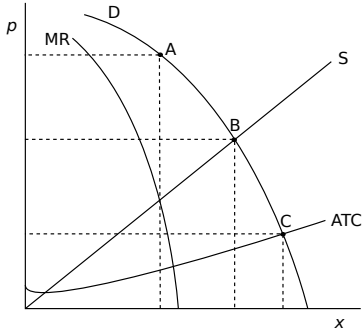}
	\caption{Demand (D, users' aggregate marginal utility), 
	supply (S, aggregator's marginal cost), 
	average total cost (ATC) and 
	marginal revenue (MR) curves 
	in the market generated by the SCPA proxy phase. 
	The segment of D between points \textsf{A} and \textsf{C} constitutes the core of the market.}
	\label{fig:CoreCE}
 \end{figure}

In order to give some intuition for the proxy auction, 
 Figure~\ref{fig:CoreCE} illustrates the market generated by the proxy phase. 
In it, the curve \textsf{D} represents aggregate demand for each price level. 
Note that aggregate demand across all slots depends on the prices in all time--slots, 
 because the users values are assumed to have inter-temporal couplings.
Therefore, the \textsf{D} shown is actually a function of $p_h$ only, with all other $\p_{-h}$ held constant. 
Figure~\ref{fig:CoreCE} also shows the aggregator's marginal revenue, supply (marginal cost) and average total cost (ATC) curves, 
and various equilibria are marked (A, B, C).

In Figure~\ref{fig:CoreCE}, the \emph{Pareto efficient} outcomes 
(in which no player can be made better off without making at least one other worse off) 
of the market fall along the demand curve between the points \textsf{A} and \textsf{C}.
The seller's profit is maximised at \textsf{A}, 
which is the rational monopolist's level of production 
(i.e.~a single producer facing an aggregate demand function composed of infinitesimally small buyers), 
and corresponds to a production level $x$ that equates the seller's marginal revenue and its marginal cost.
Neither seller nor any of the buyers benefits from 
increasing the price above this level or reducing the level of production further.

Moving down the curve, the point \textsf{B} is the \emph{perfect competitive equilibrium}, 
which is used in the case of a supply function given aggregated from the supply of a very large number producers, 
all with infinitesimally small market power, facing a similarly large number of buyers.

Finally, the point \textsf{C} is the \emph{minimum competitive equilibrium} (MCE), 
which corresponds to a price that equates the sellers' average total cost to the buyers' aggregate demand function, 
so that all of the surplus from trade is captured by the buyers.
In our setting, with only one seller, 
the MCE is the Pareto efficient outcome that minimises the aggregator's utility:
 \begin{align}
     \max_{\d}\quad	& u^i({d^i},p) 	= \max_{\d} \left[v^i({d^i}) - p^\top \cdot {d^i}, 0 \right] \quad \forall i\in\I\setminus 0 \\
      s.t.\quad 	& u^0(x,p) = p^\top \cdot x - c(x) = 0
 \end{align}
where in the second expression $x = \one\cdot \d$, 
which implies that the aggregator receives payments equal to its costs.
Any point on \textsf{D} below \textsf{C} has the aggregator trading at a loss, 
in which case it is better to not prticipate in the market at all.

At this point, the seller's profit is zero, 
and any further reductions in price or increases in production 
make it unprofitable for the seller to partici	pate in the market.

We now wish to show some properties of the proxy auction, 
that link the rules defining the final allocation and prices to the bidders' behaviour, 
under a mild assumption on their preferences.
To begin, we require the following characterisation of an auction's  outcome.

Next, we argue that the worth function satisfies \emph{bidder submodularity} (BSM), 
which is necessary and sufficient for subsequent results.
\begin{definition}\label{def:BSM}
 A worth function is bidder submodular if:
\begin{equation*}
 w(S) - w(S\setminus T) \geq \sum_{i\in T} w(S) - w(S\setminus i)\quad \forall i\in T,\ \forall T\subseteq \I.
\end{equation*}
\end{definition}

Intuitively, if $w(S)$ satisfies BSM, an additional bidder is more valuable when the market is small than when it is large.
The SCPA proxy phase is an auction for a single time--slot, 
with prices for substitute and complimentary time--slots partially determined in the preceding clock phase.
Now, residential users of electricity are known to value the first portion of their electricity use greatly,
as illustrated by an estimated own-price elasticity for electricity that are very close to zero 
(see, e.g.~\cite{FanHyndman2011}, and the survey of results therein).
Although part of this is driven by users' technological inflexibility, 
we argue that bidders' demand for electricity in a single time--slot generates a worth function that satisfies BSM, 
\emph{ceteris paribus}.\footnote{
In combinatorial domains with \emph{package bids} (bids specified over bundles of goods), 
it is quite possible that bidders violate this condition:
Imagine a bidder $i$ with high demand for a particular item that is complementary to several other bidders' packages.
Including $i$ could drive demand for the complementary goods down sufficiently far that the worth of any coalition containing $i$ 
is below that of those not containing $i$.
However, the proxy phase of the SCPA is not a combinatorial domain, as only one slot is cleared at a time
so it is difficult to construct an example violating BSM.}

\begin{theorem}
 Let the coalition worth $w(S)$ satisfy BSM,
 and assume that the aggregator computes minimal competitive equilibrium prices.
 Then the proxy phase of the SCPA implements the (VCG) allocation of the auction, 
 and straightforward bidding by the bidders is a Bayes-Nash equilibrium.
\end{theorem}

\emph{Proof sketch.}\quad
The proof has three steps, and largely follows the sequence of arguments put forward in \cite{AusubelMilgrom2002} 
for ascending price proxy auctions.  
The main difference in our setting is that the proxy auction is run implicitly by asking for a demand schedule, 
which can be interpreted as a proxy--bidder for the users, 
and (numerically) solving for the MCE prices given an aggregation of the bidders' demand schedules.

First, we have already argued that the worth function satisfies a \emph{bidder submodularity} (BSM) condition, 
as indicated by statistical evidence.

Second, call a core outcome \emph{bidder optimal} for $i\neq 0$ if it maximises $i$'s payoff over all core outcomes.
BSM implies that there is a unique bidder--optimal core outcome, 
that is unanimously preferred by all bidders.
Moreover, this core outcome corresponds to the MCE prices and allocations~\cite{BikhchandriOstroy2002}, 
and these are sufficient to support the Vickrey--Clarke--Groves (VCG) allocation of the auction 
(\cite{AusubelMilgrom2002}, Theorem~7).

Finally, implementing the VCG allocation in this way brings sincere bidding into Bayes-Nash equilibrium 
(\cite{AusubelMilgrom2002}, Theorem~8).
%
Thus, under a mild condition on the bidders' utilities and assuming 
that the aggregator computes minimal competitive equilibrium prices,
straightforward bidding is an equilibrium action in the proxy auction phase.

\subsection{The Role of the Clock Phase}
The clock phase is effectively a non-cooperative coordination game, 
in which the bidders negotiate over a price vector \p\ and a collection of (partially) coordinated demands \d\
to maximise their own utilities.
The fact that the aggregator is in control of the price adjustments 
removes the ability for bidders to engage in the most gratuitous gaming actions, 
while the RP constraint further reduces the scope for manipulations.
Thus, increasing the price of those time--slots that have a revenue deficit 
sends a clear signal to the users that these slots are more heavily congested. 
Specifically, the price discovery done by the clock phase 
allows the costs of any inter-temporal complementarities to be largely priced into each agent's decision.
As such, in the proxy phase, participants' demand functions only need to be specified over a relatively small interval,
given that the prices are partially determined for all other time--slots;
effectively, the agents are able to state their marginal utilities for electricity in the next time--slot \emph{ceteris paribus}.



\subsection{The Allocation and Price Sequence}
\noindent
In this section 
we argue that if the clock phases do manage to coordinate users' energy use over the decision horizon, 
then the sequence of allocations and prices generated by a SCPA is, in some sense, an efficient one.
We leave this as an informal argument, 
because the assumption on the efficiency of the clock phases is very strong.

On-line problems have no true optimality conditions,
and the problem to which we apply the SCPA mechanism is no different.
Given this, we highlight the following quality of the SCPA.
To begin, fix the decision horizon \H.
Then note that a one-shot clock--proxy auction, such as the one described in \cite{AusubelCM2006},
implements bidder-optimal core allocation 
(assuming BSM worth functions).
Therefore, if the SCPA were reduced to a one-shot clock--proxy auction on \H, 
then the same conditions on the outcome as above would be expected.
Next, if the clock phases do efficiently coordinate users' energy use over \H\, 
then SCPA's allocation at time--slot $t+1$ is consistent with the bidder-optimal core allocation over all \H.
The argument above can be applied recursively, 
so the allocation for all time--slots in \H\ is in the bidder-optimal core. 
Thus, if the SCPA was finite and the clock auction fully informative, 
it would sequentially implement a bidder-optimal core allocation.

There is one major caveat to this result: 
The overall efficiency of the SCPA depends greatly on the clock phase 
providing enough information to allow the households to undertake efficient scheduling of their energy use over the longer-term.
Without that, the efficiency of the proxy phases are of little merit.
Thus, identifying conditions where the clock phase is sufficiently informative is a particularly important focus of our future work.


\section{Demonstration} 
\label{sec:Demonstration}\noindent

In this section we demonstrate the SCPA.
First, we focus on the clock phase and show how prices are initially  discovered for the slots in a particular decision horizon.
Second, we show how the proxy phase operates given the prices from the preceding clock phase.
Finally, we evaluate the long-term performance of the SCPA when it used as a receding-horizon on--line mechanism.
The SCPA is a very general model that can accommodate any upward sloping supply schedule 
and a very general class of utility models --- quasi-linear utilities satisfying BSM.

\arch{
Before describing the demonstration and presenting the results, 
a brief discussion of this section's significance is warranted.
The revenue deficit and price adjustment rules used by the aggregator are trivial to implement for any monotonic cost function.
This includes those generated by piece--wise linear supply function bidding, which is an bid format used in many wholesale electricity dispatch auctions.
Note the (quadratic) cost and (linear) ATC functions that are used here are employed for demonstration purposes only, 
and as such, are kept simple for clarity and are not intended to represent the costs of participating in a real energy market.
On the other hand, most of the computation in a SCPA is carried out by users (or their EMS agents).
Thus, assuming that the utility models descibed below are appropriate for some portion of electricity users, 
general insights can be gained regarding the computation and communication requirements of a SL--DR scheme employing a SCPA. 
}

\subsection{Scenario Parameters}
We assume the aggregator faces a quadratic cost function:\footnote{
Linear and constant terms are dominated by the quadratic term, so are ignored for simplicity.}
\begin{equation*}
 c_h(x_h) = a \;{x_h}^2, 
\end{equation*}
where $a$ captures the thermal generators' efficiency, and take values of 0.002.
Furthermore, we assume that the aggregator is benevolent and only aims to only break even, 
so its supply function is given by {average total cost}: ${c_h(x_h)}/{x_h} = ax_h$.

We model a system comprising 1000 users with flexible loads, 
which all posses quasi--linear utilities as in \eqref{eqn:totalUtility}, 
but have three different types of value functions $v^i$. 
All three value functions are convex, 
but they are subject to energy use constraints of varying complexity, 
which render the overall problem non--convex.

\arch{
Each call for bids is associated with a new set of prices computed by the aggregator.
A user's response to these new prices (its \emph{bidding strategy}) could take many different forms, 
including conditioning its bid on previous levels of aggregate demand or information that it may have inferred about other bidders' preferences.
However, the analysis provided in Section~\ref{sec:Analysis} shows that the bidding and price update rules of the SCPA 
remove most of the opportunities for profitable gaming.
As such, we assume that a user employs a straightforward or truthful bidding strategy. }

\arch{
A user's straightforward bid is given by the solution to the following problem for the corresponding price level $p$:
\begin{align} \label{eqn:agentOpt}
\max_{\d^i}\quad	& v^i(d^i,\theta^i) - p^\top \cdot d^i \\
s.t.\quad		& \mbox{constraints encoded in } \theta^i. \nonumber
\end{align}
Three valuation models giving concrete specifications of $v^i(d^i,\theta^i)$ are discussed below.}

The first model of user's values for electricity is taken from \cite{SamadiEtal2012},  
which use two-piece quadratic/linear function of total electricity use over the horizon $d^i = \one^{\top}\cdot \d^i$: 
\begin{equation*}
 v^i(d^i) = 
 \begin{cases}
    \omega d^i - \frac{\alpha}{2} (d^i)^2, 
      & \mbox{ if } 0 \leq d^i < \frac{\omega^i}{\alpha^i}, \\
    \frac{{\omega^i}^2}{2\alpha^i} 
      & \mbox{ if } \frac{\omega^i}{\alpha^i} \leq d^i,
 \end{cases}
\end{equation*}
where $\alpha^i$ and $\omega^i$ are idiosyncratic parameters that determine a household's value.
Specifically, both are generated at random from a normal distribution, with
$\alpha^i \sim \mathcal{N}(0.1 , 0.02^2)$ and $\omega^i \sim \mathcal{N}(0.5 , 0.02^2)$.
In addition, energy use variables are subject to minimum and maximum constraints, 
which vary between 0 and 4kWh per slot. 
\arch{
In this model, $\theta^i$ comprises these constraints and the parameters $\alpha^i$ and $\omega^i$.
In this case,~\eqref{eqn:agentOpt} is a convex optimsation problem. }

The second model use a logarithmic function of electricity use in each slot independently: 
\begin{equation*}
 v^i(d^i) = \sum_{h\in\H} \max \left[ \alpha^i \log d^i_h,\: 0 \right],
\end{equation*}
where $\alpha^i\sim\mathcal{N}(3.0, 0.1)$ is an idiosyncratic parameter (specifically, its mean value; see below),
and energy use variables are subject to the same minimum and maximum constraints as in the first model.
\arch{
Like the first model,~\eqref{eqn:agentOpt} is a convex optimsation problem, 
and $\theta^i$ comprises these constraints and the parameter $\alpha^i$.}

The third model is taken from \cite{MhannaVerbicChapmam2015} and similar to the one formulated in \cite{KimGiannakis2013},
in which each household agent faces a cost minimisation appliance scheduling problem. 
In this, a household's value function is assumed to be constant (and large), so that it can be ignored in the optimisation problem, 
\arch{which reduces the optimisation problem in~\eqref{eqn:agentOpt} to a linear program. }
However, each appliance's energy use is subject to different constraint for different device types, 
making the problem a \emph{mixed integer} linear program (MILP). 
Specifically, the MILP's constraints include the minimum/maximum energy constraints listed above, 
more complex inter--temporal couplings, e.g. to capture devices with minimum up-times,
as well as binary variable constraints to model devices with discrete operating power points (e.g.~dishwashers and washing machines).\footnote{%
The appliances' daily energy consumption are obtained from Ausgrid's appliance usage guide, available at: \url{www.ausgrid.com.au}.}
\arch{In this model, $\theta^i$ is wholly comprised by these constraints.}

\begin{figure*}%
\centering
\begin{minipage}{0.31\textwidth}%
 \includegraphics[width=\textwidth]{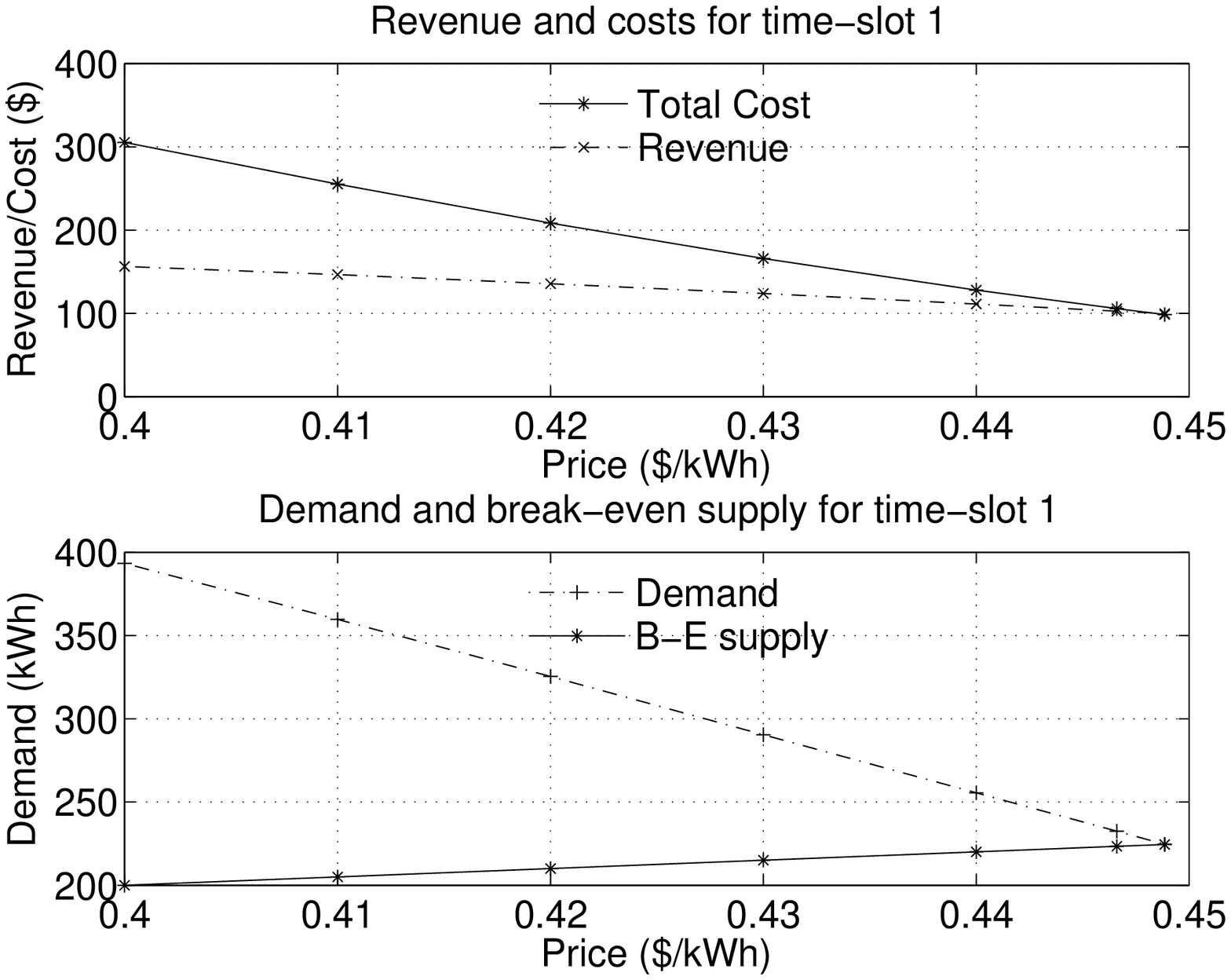}
 \caption{Clock-phase for slot $t$=1: Revenue vs.~cost, and break-even supply vs.~demand. }
 \label{fig:CP_revDef_supplyDemand}
\end{minipage}%
\quad
\begin{minipage}{0.31\textwidth}%
 \includegraphics[width=\textwidth]{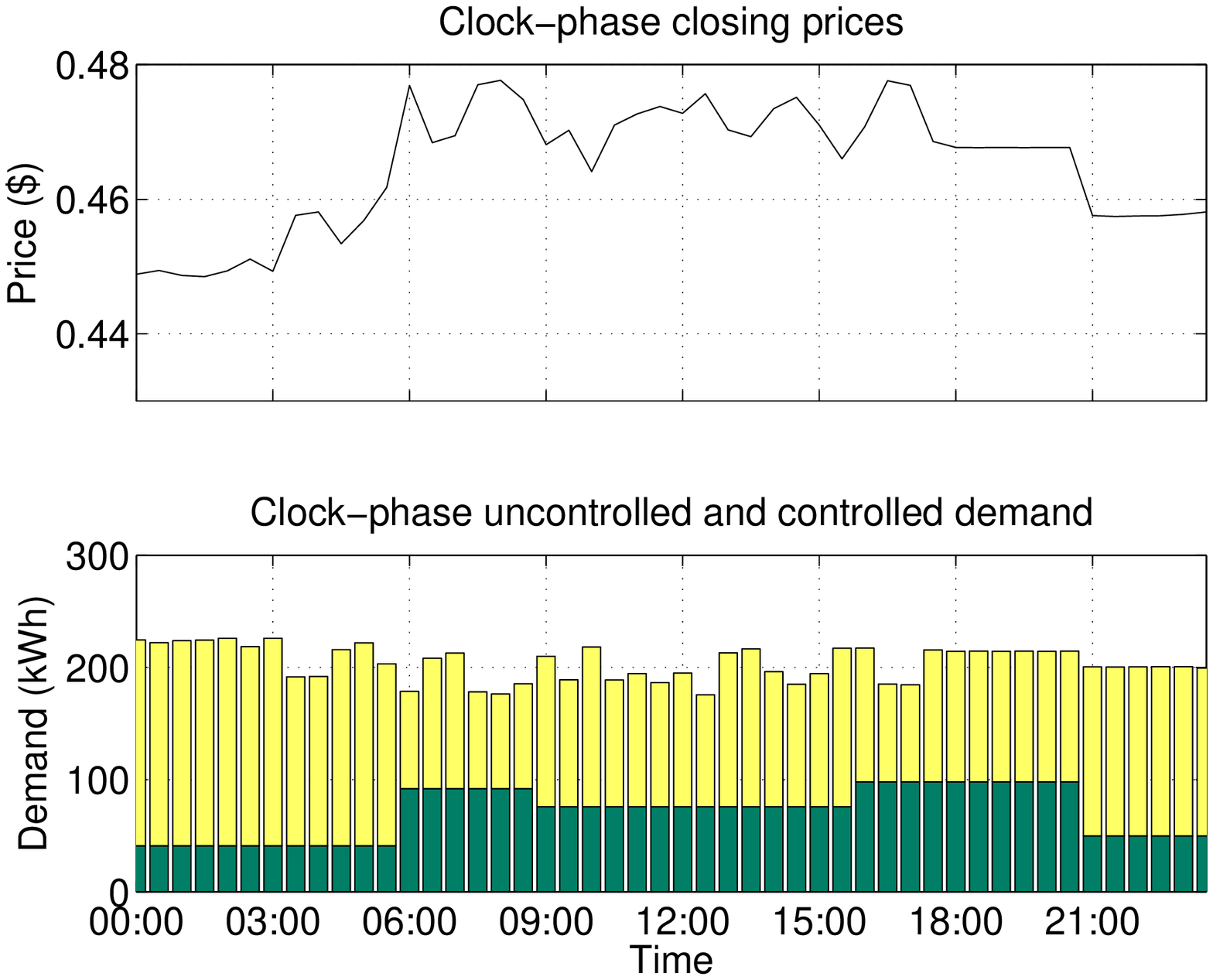}
 \caption{First ($t$=1) clock phase closing prices and demand levels for all 48 half-hourly slots.}
 \label{fig:CP_closingPriceDemand}
\end{minipage}
\quad
\begin{minipage}{0.31\textwidth}%
 \includegraphics[width=\textwidth]{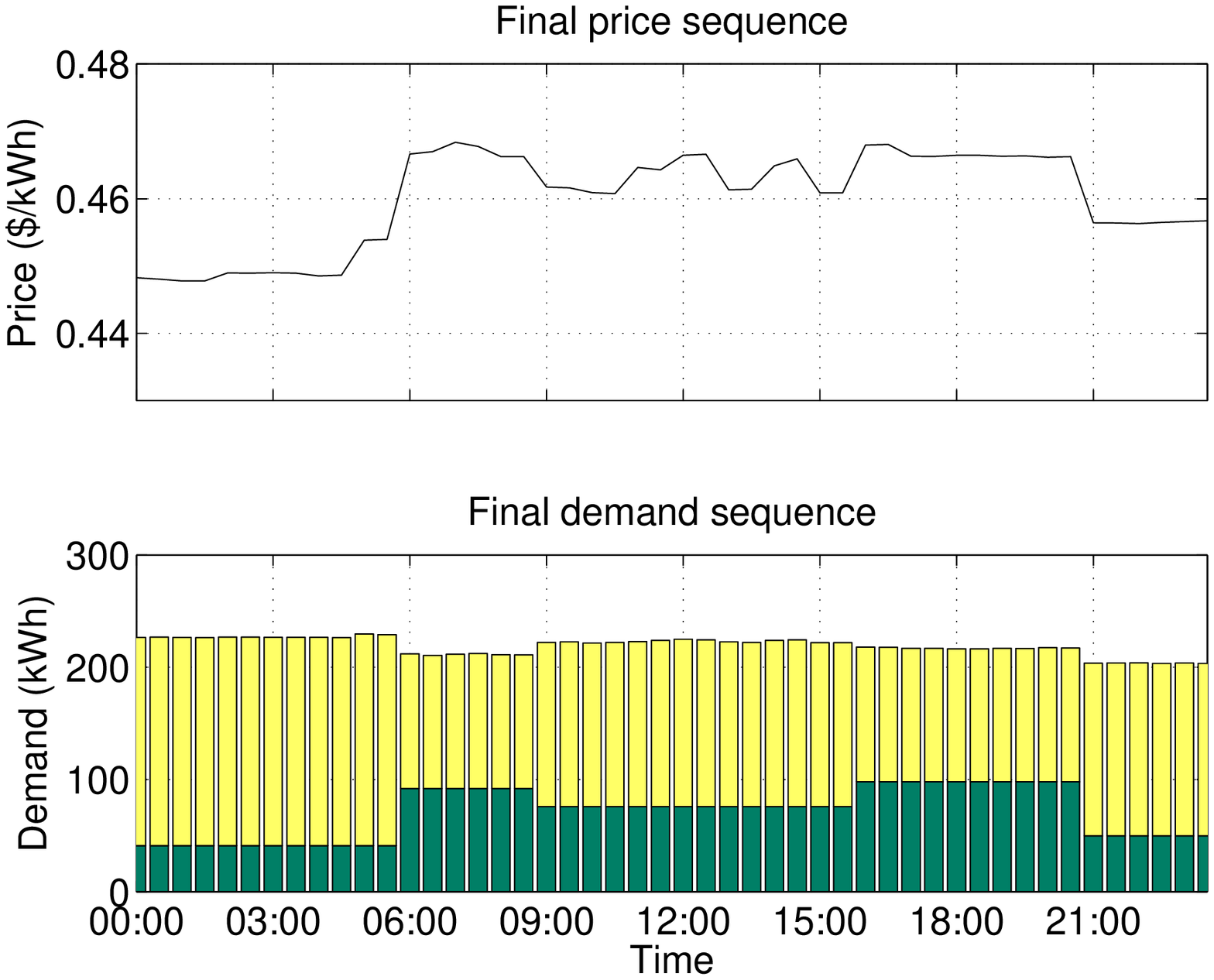}
 \caption{Final proxy-phase prices and demand levels for all 48 half-hourly slots.}
 \label{fig:FPDR}
\end{minipage}%
\end{figure*}

We also include an amount of uncontrollable load, 
which represents both energy users that are not part of the scheduling system, 
and demand that has no flexibility in its timing or magnitude 
(see the dark area in the lower panel of Fig.~\ref{fig:CP_closingPriceDemand}).

\arch{
Finally, all five components of this demonstration --- the cost function, three user preference models and uncontrolled load --- 
are subject to perturbations or noise.
Most of these perturbations are shocks to parameters and constraint values that are realised at the start of the final clock phase of time slot,
which is at most 30 minutes prior to the slot beginning.
As such, mean values are used to calculate bids and costs during the first $H-1$ rounds of the clock phase for each time--slot.
The specific shocks are described below:
The cost function parameter $a$ is perturbed my a multiplicative shock with log-normal distribution $\ln \mathcal{N}(0, 0.05^2)$ 
(i.e. with a mean of 1).
Similarly, the first and second valuation models have their upper and lower variable bounds perturbed by a multiplicative shock 
$~\ln \mathcal{N}(0, 1)$ (the same shock is applied to both upper and lower bounds), 
while the second model's valuation parameter $\alpha^i$ is also perturbed by the same degree.
The third valuation model's constraints are specified at the device level, 
and relate to the timing of use of each device; 
in contrast to the other shocks, these ``start after'' and ``end before'' constraints are randomly shifted forward or backward 
by up to three slots (1.5 hours) at the beginning of each day, 
with the shift drawn from a discrete uniform distribution. 
The uncontrolled load is randomly perturbed by a small amount of Gaussian noise (with a coefficient of variation of 0.5\%), 
which is intended to emulate unpredictable and uncontrollable loads in the system. 
}

\subsection{Initial Clock Phase Demonstration}
The clock phase operates as follows: 
For slots with a revenue deficit, price increments are computed by first
finding the zero of the line passing through the previous two increments' revenue deficits;
call this $p_{0}(k,k-1)$.  
This price indicates the point at which revenue exceeds demand 
if the $k^{th}$ iteration's electricity demand $x_h^k$ is maintained in iteration $k+1$. 
Then $\delta_h^{k+1} = \min \left[ p_{0}(k,k-1) - p_h^k, \: \bar{\delta}\right]$ is computed, 
which limits the price increment to a pre--specified maximum value.
Each $p_h^{k+1}$ is then updated by \eqref{eqn:PriceUpdate}.
In our test scenario, the maximum price increment is set to $\bar{\delta} =  \$0.01$.



The operation of the clock phase for the first slot is shown in Fig.~\ref{fig:CP_revDef_supplyDemand}. 
Each point plotted is the result of a price/aggregate demand bid pair for an iteration of the clock phase, 
starting from a price of \$0.40, with $p$ increasing at each iteration.

The upper panel of Fig.~\ref{fig:CP_revDef_supplyDemand} shows that revenue decreases as the time--slot price increases, 
as the effect of less electricity sold is greater than that of the price increase, 
but it decreases at a slower rate than total cost.
This demonstrates that the assumption on the slopes of the total cost and aggregate utility functions in Section~\ref{sec:PAanalysis} is valid.
Since we consider relatively large numbers of users, 
total cost and demand decrease smoothly with increasing price to 
a closing price of about \$0.4488 in 7 iterations, at which point the proxy auction for the first phase halted, 
although prices continued to rise for other slots until the entire proxy phase closed after 10 iterations.
Note that the maximum step size is used until the price reaches \$0.44, 
at which point the smaller step size described above is employed to reduce the overshoot of the price update step.
%
%
Finally, close observation reveals that the demand function is increasingly steep, 
validating our claim of a BSM worth function in earlier analysis.

Fig.~\ref{fig:CP_closingPriceDemand} shows the closing prices of the 48 clock auctions run in the first phase, 
and their corresponding energy allocations.
What can be most clearly seen from this is the effects of inflexible demand (the dark part of the columns) on the price.  
Those slots with the largest inflexible loads have the highest prices, 
and conversely, a large drop in price occurs at 21:00 when the uncontrolled load decreases abruptly, 
even though flexible loads move to exploit the lower prices after this time.


\subsection{Proxy Phase Demonstration}
The proxy phase was constructed by 
first, taking the final two clock phase price points to define the price interval, 
then asking for bids for each of five evenly-spaced price levels across this interval, 
and using linear interpolation to to calculate the partial equilibrium.
Here, the efficacy of our price increment rule in limiting the price overshoot is clearly demonstrated 
by the small size of the proxy-phase interval. 	

The proxy phase 
finds the MCE price to be \$0.4483 and the corresponding predicted demand to be 185.5kWh, 
with revenue and total cost equal at \$100.47. 



\subsection{Day--long Demonstration}

The day-long demonstration involves restarting the clock phase 
and opening a new clock auction for slot $t+H$ (=49) after the proxy phase closes.
In order to allow users to adjust their demands in response to random events affecting their electricity demand, 
some flexibility is injected into the restarted clock phase prices, 
in the form of price restart discounts.  
These discounts are applied to the previous clock phase closing prices to determine the opening prices in the new clock phase, 
and they are a free parameter of the auction design.
In this demonstration, we choose to set this to a uniform discount of 25\%, 
but it is our conjecture that these discounts can be tuned to provide better convergence performance.



Fig.~\ref{fig:FPDR} shows the sequence of proxy phase outcomes over the duration of one day.
Price and demand are negatively correlated, although not exactly, 
reflecting the users' preferences and constraints of the time at which they use electricity.
These are caused by a combination of:
(i) the users' constraints on electricity demand 
(use-time constraints for certain controllable devices, uncontrollable loads as in Fig.~\ref{fig:CP_closingPriceDemand}, etc.), and,
(ii) users responding to different prices at different times.
Specifically, there is high peak in price during the morning (6:00-8:30) corresponding to a peak in uncontrolled load, 
followed by a series of slightly lower peak in flexible use through the middle of the day associated with price spikes. 
In the evening (16:00-20:30) higher uncontrollable load levels again drive up the price, 
but flexible loads are able to move into late time--slots to avoid the consequent higher prices (i.e. after 21:00).
This demonstrates how the SCPA balances the dynamics of users' willingness to pay for electricity at certain times of the day 
with the cost of supply, by allowing loads to be adapted to conditions in a flexible, on--line manner.

\begin{figure}
 \centering
 \includegraphics[width=0.41\textwidth]{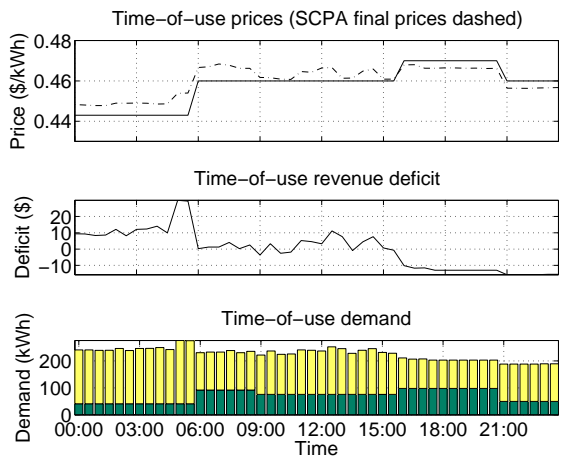}
 \caption{ToU benchmark prices, revenue deficit and demand levels.}
 \label{fig:compare}
\end{figure}

Finally, we compare the outcome of the SCPA to that of a fixed time--of--use (ToU) tariff.
The prices used in the ToU tariff are chosen heuristically to moderate the electricity use customers, 
subject to the constraint that all costs are recovered over the day 
(although in practice this condition is applied over a duration of months or even a year).  
In addition, only a few 
different prices are used, as is commonplace with existing retail ToU tariffs;
in this case three price levels are employed, with values of $(0.445, 0.460, 0.468)$ 
for time--slots 22:00-6:00, 06:00-16:30 and 20:00-22:00, and 16:30-20:00, respectively. 

Fig.~\ref{fig:compare} shows the load profile and revenue deficits induced by this tariff.  
Although the load is largely levelised across the day, 
at certain times it fails to reflect the actual costs of supplying electricity.
In particular, the spike in load between 5:00-6:00 drives the price of generation far above the ToU tariff price for that time, 
resulting in a large revenue deficit. 
More longer imbalances between price/revenue and generation cost are seen through the middle of the day, 
where cost exceed revenue, and in the evening, when revenue exceed revenue.
Indeed, because the exact timing and magnitude of these imbalances is effectively a random variable, 
such a spike in load would be seen under any fixed electricity tariff. 
In contrast, the SCPA provides a mechanism for providing energy users with timely price information 
that reflects actual electricity supply costs, demand levels, and system conditions.

%

\subsection{Computational requirements}
The demonstration above was computed using MATLAB{\reg} 
on an Intel{\reg} i7-2600 8 core CPU (3.40GHz) with 16GB of memory, 
with the agent optimisation routines run in parallel. 
%
With this setup, one day's allocation was simulated in about 97 minutes. 
It took an average of 11.6 iterations per clock phase, and with 5 price levels in each proxy phase, 
this results in $\approx$7.3s per iteration, or an average of 0.058s per agent (the aggregator's computation is trivial).
Given that a real deployment of the SCPA would run each agent optimisation routine completely in parallel, 
these computation times make an iterative auction format for on--line scheduling, 
at half-hourly or shorter intervals, entirely feasible.

\section{Conclusions and Future Work} 
\label{sec:Conclusions}\noindent
This paper develops a two-phase iterative auction mechanism 
for allocating a divisible and continually produced good (electricity) over time and on--line, 
to users that have preferences specified over demand levels in combinations of time--slots.
This mechanism is for use by an aggregator of small-load demand--response resources, 
and in this context, its design addresses three key challenges of 
combinatorial preferences, 
private information 
and scalability.

The main aim of our future work is to integrate the SCPA technique into a load-side control system comprising 
a sophisticated home energy management unit alongside control and optimisation routines for different time-scales.




\section*{Acknowledgement}
The authors acknowledge the financial contribution of the ARC and Ausgrid under the grant LP110200784.

\bibliographystyle{IEEEtran}
\bibliography{demandSideManagement}

\end{document}